\documentclass[pra,twocolumn,superscriptaddress,aps,showpacs,amsmath,amssymb]{revtex4}
\usepackage{graphicx}
\usepackage{dcolumn}
\usepackage{bm}
\usepackage{verbatim}
\usepackage{hyperref}
\hypersetup{colorlinks=true,
linkcolor=blue,
citecolor=blue,
urlcolor=blue,
filecolor=blue
}

\begin{document}

\title{Shortcut to adiabaticity in spinor condensates}

\author{Arnau Sala}
\affiliation{Departament de F\'isica Qu\`antica i Astrof\'isica, Facultat de F\'isica,\\
Universitat de Barcelona, E--08028 Barcelona, Spain}

\author{David L\'{o}pez N\'{u}\~{n}ez}
\affiliation{Departament de F\'isica Qu\`antica i Astrof\'isica, Facultat de F\'isica,\\
Universitat de Barcelona, E--08028 Barcelona, Spain}

\author{Joan Martorell}
\affiliation{Departament de F\'isica Qu\`antica i Astrof\'isica, Facultat de F\'isica,\\
Universitat de Barcelona, E--08028 Barcelona, Spain}

\author{Luigi De Sarlo}
\altaffiliation{Current address: SYRTE, Observatoire de Paris, LNE, CNRS, UPMC, 61, 
avenue de l'Observatoire, 75014 Paris, France }
\affiliation{Laboratoire Kastler Brossel, Coll\`{e}ge de France, CNRS, ENS-PSL Research University,\\
UPMC-Sorbonne Universit\'es, 11 place Marcelin Berthelot, 75005 Paris}

\author{Tilman Zibold}
\altaffiliation{Current address: Department of Physics, University of Basel, 
Klingelbergstrasse 82, 4056 Basel, Switzerland}
\affiliation{Laboratoire Kastler Brossel, Coll\`{e}ge de France, CNRS, ENS-PSL Research University,\\
UPMC-Sorbonne Universit\'es, 11 place Marcelin Berthelot, 75005 Paris}

\author{Fabrice Gerbier}
\affiliation{Laboratoire Kastler Brossel, Coll\`{e}ge de France, CNRS, ENS-PSL Research University,\\
UPMC-Sorbonne Universit\'es, 11 place Marcelin Berthelot, 75005 Paris}

\author{Artur Polls}
\affiliation{Departament de F\'isica Qu\`antica i Astrof\'isica, Facultat de F\'isica,\\
Universitat de Barcelona, E--08028 Barcelona, Spain}
\affiliation{Institut de Ci\`encies del Cosmos, Universitat de Barcelona, ICC-UB, Mart\'i i
Franqu\`es 1, E–08028 Barcelona, Spain}

\author{Bruno Juli\'a-D\'iaz}
\affiliation{Departament de F\'isica Qu\`antica i Astrof\'isica, Facultat de F\'isica,\\
Universitat de Barcelona, E--08028 Barcelona, Spain}
\affiliation{Institut de Ci\`encies del Cosmos, Universitat de Barcelona, ICC-UB, Mart\'i i
Franqu\`es 1, E–08028 Barcelona, Spain}
\affiliation{ICFO-Institut de Ci\`encies Fot\`oniques, The Barcelona Institute of Science 
and Technology,\\ Castelldefels 08860, Spain}

\date{\today}

\begin{abstract}
We devise a method to shortcut the adiabatic evolution of a spin-1 Bose 
gas with an external magnetic field as the control parameter. An initial 
many-body state with almost all bosons populating the Zeeman sublevel $m=0$, 
is evolved to a final state very close to a macroscopic spin-singlet condensate, 
a fragmented state with three macroscopically occupied Zeeman states. The 
shortcut protocol, obtained by an approximate mapping to a harmonic oscillator 
Hamiltonian, is compared to linear and exponential variations of the control 
parameter. We find a dramatic speedup of the dynamics when using the shortcut 
protocol. 
\end{abstract}

\maketitle

\section{Introduction}

Ultracold spinor Bose gases provide a beautiful example to study fragmented 
Bose-Einstein condensates (BEC)~\cite{Mue}, where Bose-Einstein condensation 
occurs in two or more single particle states simultaneously. This is an 
unusual scenario, in contrast with conventional Bose-Einstein condensation 
where bosons cluster together into a single state. For single-component 
bosons, condensation in a single state is enforced by repulsive interactions: The energetic cost 
of fragmentation is too high because of the associated exchange energy~\cite{noz1995}. 

For bosons with an internal degree of freedom, one can escape this mechanism by 
building correlations between the particles to cancel the exchange energy~\cite{Mue}. 
A spin-1 BEC with antiferromagnetic interactions in a tight trap has been predicted to host such 
fragmented condensates for vanishing magnetic 
fields~\cite{law1998a,ho2000a,koashi2000a,castin2001a,zhou2003a,barnett2010a,Ger1,Ger2}.  
The atoms condense into a single spatial mode but there remains a 
large internal degeneracy at the single-particle level. Antiferromagnetic 
interactions lift this degeneracy, and lead to a total spin-singlet ground 
state which is completely fragmented between the three sublevels. The 
many-body singlet state displays strong quantum correlations, and has attracted 
much theoretical interest. 

This spin-singlet fragmented condensate is fragile against any perturbation lifting 
the single-particle degeneracy, such as external magnetic 
fields~\cite{ho2000a,koashi2000a,castin2001a}. In experiments with alkali 
atoms, the most relevant perturbation is the quadratic Zeeman splitting 
between the Zeeman sublevels $m=0$ and $m=\pm 1$~\cite{stamperkurn2013a}.
For finite atom number $N$, there is a small but non-vanishing window where 
the singlet state survives as the quadratic Zeeman splitting increases 
from zero, before a crossover to a single $m=0$ condensate takes place 
(``single BEC domain'')~\cite{barnett2010a,Ger1,Ger2}. An appropriate 
witness of the transition is the variance 
$\Delta N_0\equiv \sqrt{\langle N_0^2\rangle -\langle N_0\rangle^2}$ 
which goes from $\propto N$ in the spin-singlet state to 
$\propto \sqrt{N}$ in an uncorrelated many-body state~\cite{Ger2}.  

\begin{figure}
\includegraphics[width=\linewidth]{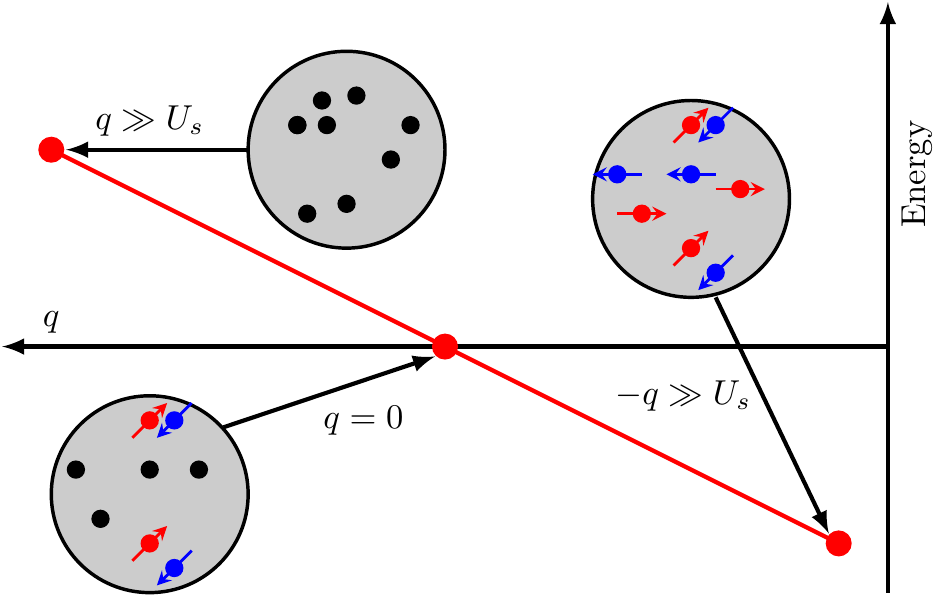}
\caption{Sketch of the proposed experimental protocol. A spin-1 BEC is prepared at large 
positive values of the quadratic Zeeman energy (QZE) $q$ with all atoms in the $m=0$ state. 
In this regime, the initial state is very close to the ground state (arbitrarily close 
as $q\rightarrow\infty$). In the adiabatic method, the QZE is slowly reduced in such 
a way that the state of the system remains always close to the instantaneous ground state. 
Stopping the QZE ramp at $q=0$, the system ends up in a total spin-singlet ground state 
with strong spin correlations. Stopping the ramp at a large, negative value of $q$, we 
prepare instead a twin Fock state with half the atoms in the Zeeman $m=\pm1$. In 
this paper we target the production of the singlet state, and examine this procedure 
and alternative ramps which are not adiabatic but result in a state close to the ground 
state in a much faster time. \label{fig:rampq}}
\end{figure}

Because of the sensitivity to external perturbations, the singlet state has so 
far eluded experimental observation. The gap to the first excited states is low 
and scales as the inverse of the number of atoms~\cite{law1998a}. Evaporative cooling 
used to produce quantum gases is unable to reach such ultralow temperatures. 
Another procedure is to adiabatically produce the singlet state by first applying 
a magnetic field and condensing in the $m=0$ state, and then slowly remove the 
field to produce the desired singlet state --- see the sketch in Fig.~\ref{fig:rampq}. 
In order to stay adiabatic, the dynamics must be very slow in view of the small 
energy scales involved, making the procedure vulnerable to heating or inelastic losses.

In this article we introduce a way to shortcut the adiabatic following 
and thus produce the desired final state in times much shorter than 
those needed in adiabatic processes. Such methods have been recently 
derived for a number of quantum mechanical systems --- see for instance 
Ref.~\cite{torr12}, and promise to provide important advances in actual 
implementations of quantum technologies, for instance trapped ions~\cite{adc16}. 
Exact protocols have been derived for particular problems, e.g. the quantum 
harmonic oscillator~\cite{muga}. In other cases, approximate procedures, 
obtained by adapting exact ones, have been proven to be quite promising 
when applied to quantum many-body systems~\cite{Bru1,Bru2,sc15}. 

As will be shown, the approximate shortcut protocol will be obtained from a 
large $N$ limit of the quantum many-body system. This limit will allow us to 
map our original many-spin problem into an effective harmonic oscillator, for 
which an exact solution is available~\cite{muga}. Interestingly, very recently 
a similar harmonic description of a spinor BEC has allowed the authors in Ref.~\cite{chap} 
to prove parametric amplification of a spinor system. This work proves experimentally 
the appropriateness of the harmonic description.

The article is organized as follows. In Section~\ref{sec2}, we present 
the theoretical model to describe the spinor BEC and discuss 
the adiabatic preparation of the ground state. In Section~\ref{sec:sa} we 
obtain our protocol to shortcut the adiabatic evolution in the spinor system 
from a continuum approximation to the spin dynamics. In Section~\ref{sec3} 
we apply our shortcut protocol to the BEC regime (dominated by the quadratic 
Zeeman energy). In Section~\ref{sec4} we consider a broader range of 
parameters, discussing the quality of our protocol to produce fragmented 
BEC starting from the BEC side. In Section~\ref{sec5}  we present results 
making use of current experimental setups~\cite{chap}. In Section~\ref{sec6}, we briefly 
summarize our work and present the main conclusions.

\section{Theoretical Model}
\label{sec2}
\subsection{Description of the system}

We consider an ultracold gas of spin-1 bosons in a harmonic trap under the 
action of an external magnetic field. We assume a single 
spatial mode in the trap, that is, all bosons condense in the same spatial 
orbit irrespective of their internal state. With this assumption we are left 
with three single-particle states, $|+1\rangle$, $|0\rangle$ and $|-1\rangle$, 
corresponding to the Zeeman states with magnetic quantum numbers $m=+1,0,-1$, 
respectively. The linear Zeeman effect acts only as a shift in the energy 
and does not contribute to determine the equilibrium state. The main contribution 
of the magnetic field is the quadratic, or second order, Zeeman (QZ) 
effect~\cite{stamperkurn2013a}. Under these assumptions, the system is well 
described by the Hamiltonian~\cite{barnett2010a,Ger1}
\begin{equation}\label{eq:ham1}
\hat H= \frac{U_s}{2N}\ \hat{\bm S}^2-q\hat{N}_0\, ,
\end{equation}
where $U_s>0$ is the spin interaction energy per atom, $N$ is the number of 
atoms, ${\hat{\bm S}}^2$ is the (dimensionless) total spin operator, $q$ is the 
quadratic Zeeman energy and $\hat N_m$ is the number operator of the Zeeman 
state $m=0,\pm1$.

The first term in the right-hand side of Eq.~(\ref{eq:ham1}) describes 
antiferromagnetic interactions between pairs of atoms, and favours configurations 
with low total spin $S$. In absence of the quadratic Zeeman term, $q=0$, the 
eigenstates are known analytically and are given by the total spin eigenstates 
$|N,S,M\rangle$, where $S$ is the total spin and $M$ the eigenvalue of $\hat{S}_z$, 
the projection of the total spin on 
the $z$ axis. This is the basis we will be using in the 
following sections. Low-$S$ configurations are obtained by putting many spin-1 atoms to 
form singlet state pairs, while the remaining atoms can occupy any Zeeman 
sublevel. For practical convenience, from now on $N$ will be set to an 
even number. The ground state for even $N$ is the total spin singlet  
$\vert N, S=0, M=0 \rangle$. This highly fragmented state, termed 
``spin-singlet condensate'' (SSC), takes the form of a condensate of 
delocalized spin-singlet pairs,
\begin{eqnarray}\label{SSC}
|{\rm SSC}\rangle \propto & \left( \hat{A}^\dagger \right)^{N/2}\vert \textrm{vac}
\rangle
\end{eqnarray}
where $A^\dagger=(\hat{a}_0^\dagger)^2-2\hat{a}_{+1}^\dagger \hat{a}_{-1}^\dagger$ creates a 
pair of atoms in the two-particle singlet state, $\hat{a}_m$ is an annihilation operator 
for a particle in the Zeeman state with third component of the angular momentum 
equal to $m$, and $\vert \textrm{vac}\rangle$ is the boson vacuum.
This fragmented state is characterized by three macroscopically 
populated states, $\langle N_{+1} \rangle=\langle N_{0} \rangle=
\langle N_{-1} \rangle=N/3$, with large fluctuations of the individual components~\cite{Mue}.

The second term in Eq.~(\ref{eq:ham1}) describes the interaction of the 
system with the external magnetic field. In the non-interacting limit 
$U_s \rightarrow 0$ and for $q>0$, the QZE forces all the spins to occupy the 
state $m=0$, thus forming a single Bose-Einstein condensate 
with $\langle N_0 \rangle=N$ and $\langle N_{+1} \rangle=\langle N_{-1} \rangle=0$, 
the so-called $z$-polar state,
\begin{equation}
|{\rm Polar} \rangle_z \propto \left( \hat{a}_{0}^\dagger\right)^{N} \vert \textrm{vac}\rangle \,.
\end{equation}

It is worth noting that $M$ remains fixed when changing $q$, because the 
Hamiltonian commutes with $\hat{S}_z$. This is 
a good approximation to the behavior due to the experimental conditions 
of the atomic quantum gases, which are highly isolated from the environment, 
and to the microscopic rotational invariance of the spin exchange 
interaction~\cite{stamperkurn2013a}. 

For simplicity, we take $M=0$. Also, as the number of particles $N$ is fixed during the 
evolution, we will omit it on the kets, thus, for now on we will use 
the notation $|S\rangle \equiv |N,S,0\rangle$. 

\subsection{Ground state for intermediate values of $\vert q \vert$}

For generic values of $q, U_s$, we write a general state $|\phi\rangle$ with fixed 
$N$ and $M=0$ as $|\phi\rangle = \sum_S c_S |S\rangle$. The Schr\"odinger equation 
$\hat{H} |\phi \rangle=E|\phi \rangle$ in the $S$ basis reduces to the following 
discrete eigenvalue equation (see Appendix~\ref{app:A}),
\begin{equation} \label{eq:ham2}
h_{S,S+2}\, c_{S+2} +h_{S,S-2}\, c_{S-2} + h_{S,S}\, c_S= E\, c_S\, .
\end{equation}

In the upper panel of Fig.~\ref{fig:cx} we show the transition from the 
$U_s$-dominated fragmented regime to the single BEC regime when varying the 
ratio $q N^2/U_s$. The transition between the two regimes takes place at 
values $q  N^2 U_s \simeq 1$ and is seen in the behavior of the variance 
$\Delta N_0/N$ of the populations in the $m=0$ Zeeman state. As explained 
in the introduction in the uncorrelated BEC state, $\Delta N_0 \propto \sqrt{N}$, 
while in the spin-singlet state the fluctuations are much larger, $\Delta N_0 \propto N$.

\subsection{Adiabatic preparation of the singlet ground state}
\label{XXX}

Experimentally, the value of the QZE can be controlled easily in real time. For 
instance, for Sodium atoms with hyperfine spin $F=1$ in a magnetic field $B$, 
the quadratic Zeeman shift contributes a {\it positive} amount to $q$. It is also 
possible to achieve $q<0$ by using the differential level shift induced on the 
individual Zeeman sublevels by a far off-resonant microwave field (see~\cite{gerbier2006b} 
for details). With a suitable choice of the microwave polarization, detuning and power, 
the sign and magnitude of $q$ can be changed at will.

This experimental control of the QZE opens a way to the generation of 
strongly correlated states in spin-1 quantum gases. The principle is the 
following. For zero magnetization and a large and positive QZE, the ground
state is very close to a single BEC with all atoms in the $m=0$ Zeeman state. 
A good approximation of this state can be prepared ``by hand'', \textit{e.g.} 
by applying radio-frequency ---rf--- pulses with suitable frequency and 
polarization to a spin-polarized ensemble in $m=+1$, for instance. Starting from 
this initial state and decreasing slowly the value of $q$, the system will 
adiabatically follow its ground state, and end up prepared in the SSC state given 
in Eq.~(\ref{SSC}) when $q\approx 0$. 

We can estimate the speed at which the magnetic field should be decreased by the 
usual adiabatic criterion, $\vert \langle j \vert \dot{H}\vert i \rangle \vert \ll\hbar\omega_{\rm ji}^2$, 
where $\vert i \rangle$ and $\vert j \rangle$ are two eigenstates of the 
Hamiltonian. The dangerous region is around $q< U_s/N^2$, where the energy gap 
to the first excited state takes its minimum value $\sim 3U_s/N$. In this region, 
the QZE ramp has to be very slow. We make a crude estimate by assuming that $q$ 
decreases between $U_s$ and $0$ in a time $\tau$. Also in this region, $N_{\pm 1}$ 
are on the order of $N/3$. This leads to 
$\vert \langle j \vert \dot{H}\vert i \rangle \vert\sim N U_s/3\tau$ and to 
the adiabaticity criterion,
\begin{equation}\label{tau_ad}
\tau \gg \frac{N^3\hbar}{27 U_s}.
\end{equation}
The catastrophic scaling $\tau \propto N^3$ shows that this method will be limited 
to small, mesoscopic samples. Using very long ramp times to fulfill the adiabaticity 
criterion will make the protocol vulnerable to experimental limitations not captured 
by the single-mode Hamiltonian, such as technical heating (specific for each experimental 
setup) and inelastic losses (specific for each atom).

Inelastic atom losses destroy the rotational symmetry since atoms are lost 
at random from any Zeeman state. A common source of inelastic losses is 
three-body recombination into a weakly-bound molecule and a fast atom, 
resulting in three atoms lost from the trap. The total rate of these events 
can be written as $N\Gamma_3$, where $\Gamma_3 = (K_{3B}/N)\int d^3\bm{r}\, n(\bm{r})^3$ 
is determined by a species-dependent rate constant $K_{3B}$ and by the 
spatial density $n$. Demanding less than one single inelastic event 
(on average) during the entire adiabatic protocol gives a bound $1/\tau \gtrsim N\Gamma_3$. 

For illustrative purposes, we consider a gas of atoms condensing 
in the Gaussian ground state of a tight harmonic trap of frequency 
$\omega$. The Gaussian ground state of the trap is a good approximation 
of the actual condensate wavefunction for sufficiently low atom number 
$N < \sigma/\overline{a}$, with $\sigma=\sqrt{\hbar/m_{\rm A}\omega}$ the 
harmonic oscillator length, with $m_{\rm A}$ the atomic mass and with 
$\overline{a}$ the spin-independent $s-$wave scattering length. The 
spin-dependent scattering length $a_s$ is determined by the relation 
$U_s=(4\pi\hbar^2 a_s/m_{\rm A}N) \int d^3\bm{r}\, n(\bm{r})^2$~\cite{law1998a}. 
The bound $1/\tau \gtrsim N\Gamma_3$ can be written as a bound on the 
maximum affordable atom number in trap, written in compact form as
\begin{align}\label{N3}
N \ll 5.1 \left(\frac{\sigma}{l_{3B}}\right)^{3/5} ,
\end{align}
where $l_{3B}=(m_{\rm A} K_{3B}/\hbar a_s)^{1/3}$ has the dimension of a length. 

We specialize to the case of $F=1$ Sodium atoms, where $\overline{a}\approx 2.5\,$nm 
and $a_s\approx 0.1\,$nm \cite{knoop2011a} and three-body loss rate constant 
$K_{3B}\sim 1.6 \times 10^{-30}$at.cm$^{6}$/s~\cite{goerlitz2003a}. Using 
$\omega/(2\pi)=2\,$kHz, one finds $N \ll 36$ for the parameters given above, 
showing that the adiabatic approach is reserved for mesoscopic samples containing 
only a few atoms. This motivates us to find alternative solutions enabling 
a substantial speed-up of the dynamics, which is our main objective in the 
rest of this paper.

\section{Shortcuts to adiabaticity}
\label{sec:sa}

In view of the limitations of the adiabatic approach described above, we now 
examine a different method where the same final result can be reached in a 
much shorter time. In the literature, there are well-established shortcut 
protocols for one-body harmonic potentials~\cite{muga}. Our strategy is to use 
these results to manipulate the many-spin system of interest by mapping it to 
an effective harmonic oscillator problem. We show in this section how a reasonable 
harmonic approximation to the many-body problem can be derived. By means of 
such approximate equation we map the shortcut protocol to the exact time 
dependent Schr\"odinger equation built from Eq.~(\ref{eq:ham2}). 

\subsection{Continuum approximation}

The first step consists in 
deriving a continuum approximation to the Hamiltonian, Eq.~(\ref{eq:ham2}). 
For large $N$ and considering $1\ll S\ll N$, the coefficients $c_S$ can be 
assumed to vary smoothly from $S$ to $S\pm2$. Hence, $c_S$ can be approximated 
by a continuous function $c(x)$, where $x\equiv S/N$ varies from $0$ to $1$. 
Following the derivations in Appendix~\ref{app:B}, we arrive at an effective
Schr\"odinger-like equation for a harmonic oscillator
\begin{equation} \label{eq:ham3}
-\frac{\hbar^2}{2M^\ast}c''(x)+\frac{M^\ast\omega^2}{2}x^2 c(x)=(E+Nq)c(x) \, 
\end{equation}
with the oscillator frequency given by $\hbar\omega=\sqrt{q(q+2U_s)}$ and the 
oscillator ``mass'' by $M^\ast/\hbar^2=N/2q$. The ground state obeying the boundary 
condition $c(0)=0$ is the wave function
\begin{align} \label{eq:cx}
c(x)=&\frac{2 \sqrt{2}}{(\pi \sigma)^{1/4}} \frac{x}{\sigma} 
\exp{\left( -\frac{x^2}{2 \sigma^2} \right)} \\
\sigma =& \sqrt{\frac{2}{N}} \left( \frac{q}{q + 2 U_s} \right)^{1/4}\, ,
\end{align}
with energy 
\begin{align} \label{eq:en}
E=\frac{3}{2} \sqrt{q(q+2 U_s)} - Nq\, .
\end{align}

\begin{figure}
\includegraphics[width=\linewidth]{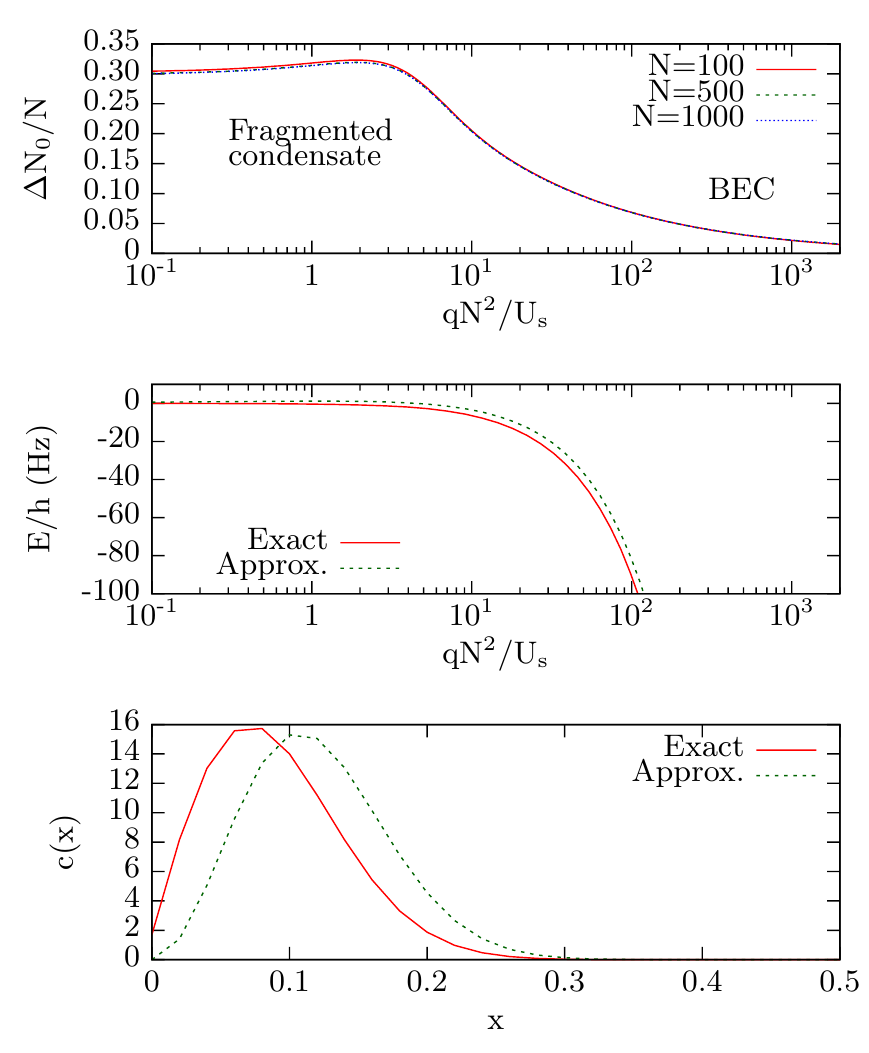}
\caption{Upper panel: Fluctuation of the number of particles in the 
$m=0$ manifold, $\Delta N_0/N$ computed for three different number 
of atoms. Middle panel: Energy of the ground state of the exact 
system compared with the energy of the ground state obtained in 
Eq.~(\ref{eq:en}) for different values of the parameter $q$. When 
$qN^2 U_s^{-1} < 2$, the system becomes $U_s$-dominated and, thus, 
the energy is constant [see Eq.~(\ref{eq:en})]. It is worth noting that 
both in the fragmented ($U_s$-dominated) and BEC ($q$-dominated) 
regimes, the approximate value of the energy agrees well with the exact one. 
Lower panel: Wave function of the ground state compared to the one obtained 
with the continuous approximation described in the text. The middle and lower 
panels are obtained for a system of $N=100$ spins. In all cases, 
$U_s/h = 104.13$ Hz. For the lower panel, we have used $q=U_s$. \label{fig:cx}}
\end{figure}

In Fig.~\ref{fig:cx} we compare the approximated and the exact solutions 
of our Hamiltonian. In the middle panel of Fig.~\ref{fig:cx} we see that 
the energy of the ground state is well reproduced by the harmonic 
approximation. In particular it is interesting to note that the harmonic 
approximation works well both in the $U_s$-dominated regime and in the 
$q$-dominated one. Comparing the actual wave functions in the lower panel of 
Fig.~\ref{fig:cx}, we can see that the solution of the approximate Hamiltonian 
has a similar shape as the exact wave function although its maximum is slightly 
displaced towards higher values of $S/N$.

\subsection{Shortcut protocol to the adiabatic evolution}

The idea behind the shortcut to adiabaticity in the time-dependent evolution of an 
harmonic oscillator is the following. First we consider that the system 
is initially in the ground state for a certain initial value $q(0)$ 
of the control parameter. Then, we impose that at a given time $t_f$ 
the system must be exactly in the ground state for a different 
value of the control parameter, $q(t_f)=q_f$. The goal is, thus, to find a 
function $q(t)$ that does the job. If the final time is sufficiently large, 
then any smooth ramp of the control parameter should work, since the 
evolution would be adiabatic. For short ramp times, an arbitrary 
ramp function would in general result in the excitation of many modes besides 
the ground state at the final time. The goal is therefore to engineer the 
ramp function in such a way as to minimize the excitations at $t=t_f$ and beyond,   
\textit{i.e.} one seeks to produce an almost stationary state once 
the ramp is completed.

The Schr\"odinger-like equation in Eq.~(\ref{eq:ham3}) is already close to 
the one corresponding to a harmonic oscillator. The control parameter is the QZE, $q=q(t)$. 
The term on the right-hand side (a shift in the total energy) does not 
have any effect on the dynamics. We also 
limit ourselves to the regime $q\ll U_s$. The final Schr\"{o}dinger-like 
equation Eq.~(\ref{eq:ham3}) is that of a harmonic oscillator 
with time-dependent ``mass'' and frequency $\omega^2(t)=2 q(t) U_s/\hbar^2$.
A similar equation was considered in Ref.~\cite{Bru2} to describe the dynamics of 
a two-mode Bose-Hubbard model. Following the same method, we look for a self-similar 
solution $c(x) =c_0(x/\rho)/\sqrt{n}e^{i{\Theta(\rho,\dot{\rho}})}$, with a scaling parameter 
$\rho$. Such a solution exists if the scaling parameter obeys the so-called Ermakov 
equation~\cite{muga}, 
\begin{align}
\ddot{\rho} + \omega(t)^2 \rho = \frac{\omega_0^2}{\rho^3}. 
\end{align}
The constant $\omega_0$ is just an integration constant that we set to 
$\omega_0=\frac{1}{\hbar}\sqrt{2 q_0 U_s}$. This, together with the substitution 
$b=1/\rho$ gives
\begin{align} \label{eq:erma}
\frac{2 \dot{b}^2}{b} - \ddot{b} + \frac{2 q(t) U_s}{\hbar^2} b 
= \frac{2 q_0 U_s}{\hbar^2} b^5.
\end{align}
$b(t)$ is an arbitrary function that only has to satisfy the frictionless conditions
\begin{align}
\displaystyle b_i\equiv &\, b(t_0) = 1 \, , \nonumber \\
\displaystyle b_f\equiv &\, b(t_f) = \left(\frac{q_f}{q_0}\right) ^{1/4} \, , \nonumber\\
\dot{b}(t_0)=&\, \dot{b}(t_f)=\ddot{b}(t_0)= \ddot{b}(t_f)=0 \, , \nonumber
\end{align}
where $t_0$ will always be zero. There is an infinite set of functions $b(t)$ 
that can be used for this purpose, as the boundary conditions provide ample 
freedom to choose $b(t)$. Here we will use a simple polynomial ansatz, taken from Ref.~\cite{muga}, 
\begin{equation}
b(t)=b_i+10(b_f-b_i)s^3-15(b_f-b_i)s^4+6(b_f-b_i)s^5\, , \label{eq:b}
\end{equation}
where $s=t/t_f$. Thus, the function $q(t)$ is
\begin{align} \label{eq:qt}
q(t) =  \left( 2 q_0 U_s b^4 + \hbar^2 \frac{\ddot{b}}{b} 
- 2 \hbar^2 \frac{\dot{b}^2}{b^2}\right) \frac{1}{2 U_s}.
\end{align}
The six frictionless conditions previously mentioned uniquely determine the 
fifth-order polynomial chosen. We have tested that using a sixth-order polynomial 
or power-law functions did not improve the results, hence Eq.~(\ref{eq:b}) is used 
in the rest of this work. Let us remark that the freedom in choosing $b(t)$ 
can be used to design more constrained protocols depending on the specific needs, 
e.g. avoiding too large values of $q(t)$.

\section{Shortcut to adiabaticity in the BEC regime}
\label{sec3}

\begin{figure*}
\includegraphics[width=\linewidth]{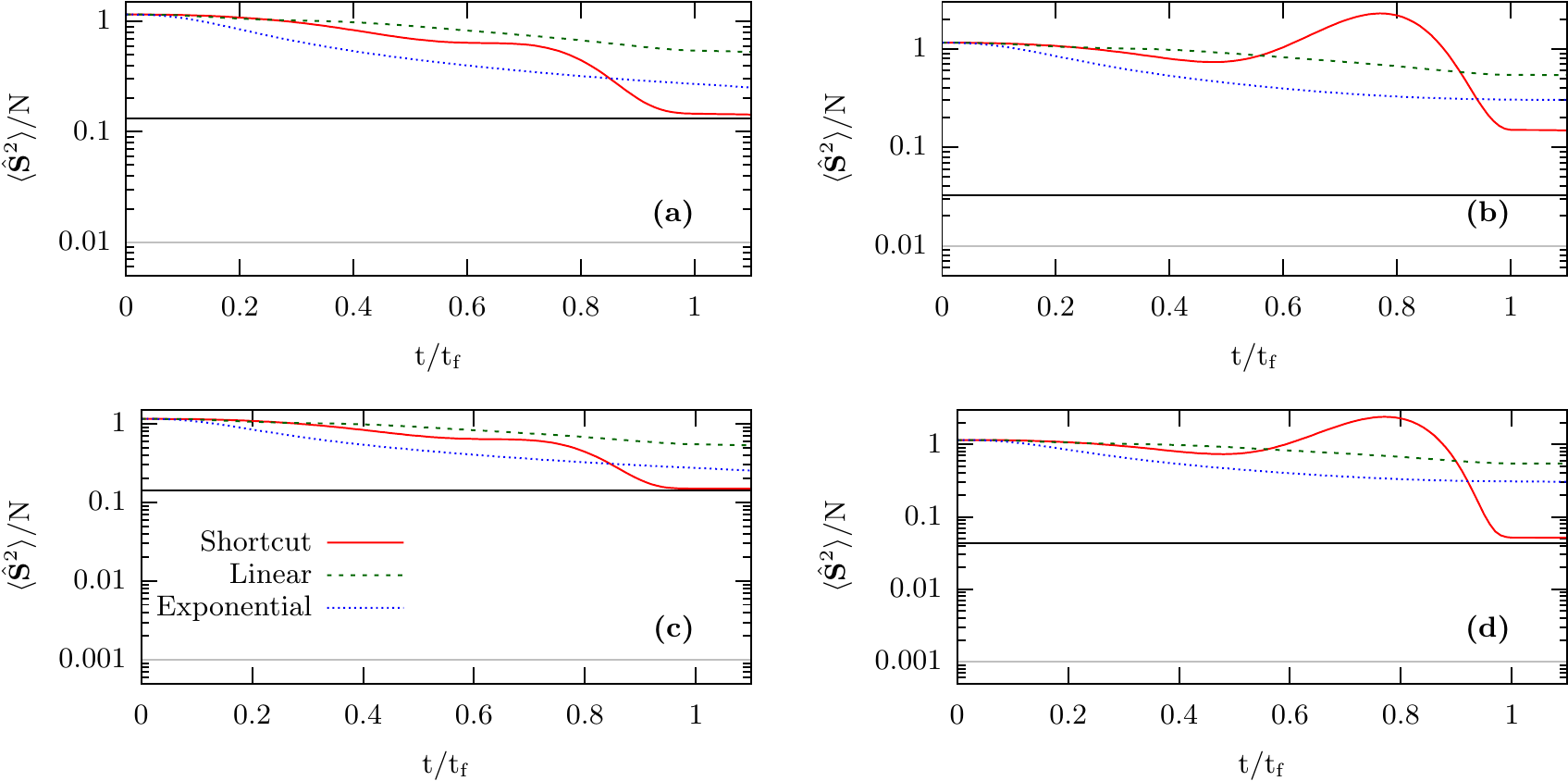}
\caption{Mean squared spin $\langle \hat{\bm S}^2 \rangle$ as a function of 
time computed for systems with $N = 100$ and $N = 1000$ spins evolved following 
either the shortcut protocol, a linear or an exponential ramps. All  
cases describe the evolution of the exact solution of the Schr\"{o}dinger 
equation in situations where the evolution is clearly not adiabatic. Only 
the shortcut protocol brings the system to a state with lower total spin $S$. 
In all panels the initial state corresponds to $N^2 q_0/U_s=N^2$, thus deep 
in the BEC phase (see Fig.~1 of~\cite{Ger1}). Panels on the left, 
---\textbf{(a)} and \textbf{(c)}--- target a state at $N^2 q_f/U_s=0.01N^2$, 
whereas right panels \textbf{(b)} and \textbf{(d)} correspond to a final 
state closer to the fragmented phase $N^2q_f/U_s=0.001 N^2$. \textbf{(a)} and \textbf{(b)} 
correspond to $N=100$ while \textbf{(c)} and \textbf{(d)} are computed with 
$N=1000$.  The horizontal solid black and light grey lines show the level 
of fluctuations in the ground state for $q=q_f$, and 
$\langle \hat{\bm S}^2 \rangle=1/N$, respectively. 
In all cases, $U_s/h = 104.13$ Hz and $t_f=0.01$~s. \label{fig:fid-s2}}
\end{figure*}

In this Section, we consider the performance of our shortcut protocol in the 
BEC regime. That is, our main goal is to evolve from the ground state 
of Eq.~(\ref{eq:ham1}) for an initial value of $q(t=0)=q_0=U_s$ to the 
corresponding ground state for $q(t_f)=q_f$ such that $q_0 > q_f \gg U_s/N^2$. 
Strictly speaking, the choice $q_0=U_s$ does not comply with the assumption 
used to derive the protocol, $q(t)\ll U_s$. The reason why the protocol can 
be extended to this situation is that for short times $t \lesssim  t_f/4$, 
the protocol produces a very small variation of $q$, and the dynamics thus 
remains mostly adiabatic.

To judge the quality of the shortcut protocol we will compare it to 
two other ramps shapes, linear and exponential,
\begin{align}
q_{\rm lin}(t)=& q_0-(q_0-q_f)\, t/t_f \, , \\
q_{\rm exp}(t)=& (q_0-q_f) \frac{e^{-\alpha t/t_f}-e^{-\alpha}}{1-e^{-\alpha}}+q_f \,.\nonumber
\end{align}
The linear ramp is uniquely determined, and the exponential ramp was found to 
provide the best results for a decay constant $\alpha= 5$, a value we have used 
in all the reported results.

We benchmark our shortcut protocol by numerically solving the full time 
dependent Schr\"odinger equation with $\hat{H}$ from Eq.~(\ref{eq:ham1}) 
for a particular ramp. We used the mean-squared spin 
$\langle \hat{\bm S}^2\rangle$ as a fidelity witness. Note that in the 
regime we consider in this Section, the target final 
value of $q_f$ is far above the value $\sim U_s/N^2$ below which the ground state 
reduces to the total spin singlet state. As a result, the value of 
$\langle \hat{\bm S}^2\rangle$ in the ground state corresponding to 
$q_f$ fulfills $1 \ll \langle \hat{\bm S}^2\rangle \ll N$. 

\begin{figure}[t]
\includegraphics[width=\linewidth]{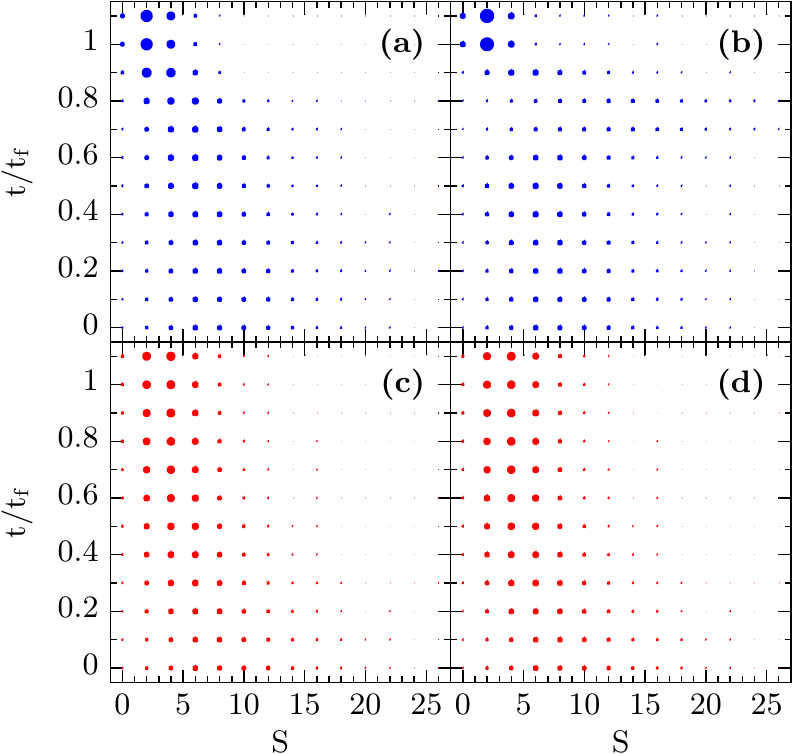}
\caption{Time evolution of the wave function. The radii of circles indicate the 
value of $|c_S|^2$ for each $S$ at each time $t/t_f$. Blue circles correspond to 
a system evolved using the shortcut protocol and red circles correspond to a system 
evolved with an exponential ramp. Panels \textbf{(a)}, \textbf{(c)} show the 
evolution of a system from $q_0/h=104.13$ Hz to $q_f=10^{-2} q_0$ in 
$t_f=0.01$ s and panels \textbf{(b)}, \textbf{(d)} show the evolution of a system 
from $q_0/h=104.13$ Hz to $q_f=10^{-3} q_0$ in $t_f=0.01$ s, so this 
figure can be directly compared to Fig.~\ref{fig:fid-s2}. Since only the even 
solution is considered, the wave function $c_S$ at odd $S$ is identically 0. 
\label{fig:cs-3}}
\end{figure}

In Fig.~\ref{fig:fid-s2} we present the first results, corresponding to two 
different dynamical situations. The first one shown in Fig.~\ref{fig:fid-s2}\,(a,c) goes from $q_0/U_s=1$ to 
$q_f/U_s=0.01$. The second one shown in Fig.~\ref{fig:fid-s2}\,(b,d) goes one order of 
magnitude smaller, to $q_f/ U_s=0.001$. Also we compare in the figure two different 
values of $N=100$ and $1000$. Several features can be observed. In all cases, the 
shortcut protocol performs clearly better than the other two ramps, while the 
exponential ramp performs better than the linear ramp. The spin witness 
$\langle \hat{\bm S}^2\rangle$ at the final time at $t_f=0.01\,$s is substantially 
lower for the shortcut protocol ($\langle \hat{\bm S}^2 \rangle$ decreases by 
an order of magnitude from its initial value), and closer to the value expected 
in the final ground state for larger $N$. In comparison, the other two protocols 
are only able to decrease it by at most a factor of 4 in the same time. Moreover, 
the final value of $\langle \hat{\bm S}^2 \rangle$ decreases with increasing 
atom number at a fixed $t_f$. Equivalently, the final fidelities obtained 
with the shortcut protocol improve as we increase $N$. This could be expected as we 
have obtained our protocol in the large $N$ limit, thus making the protocol closer to 
an exact description as $N$ is increased. The results obtained with the exponential and 
linear ramps are mostly independent of the number of particles.

\begin{figure}
\includegraphics[width=1.\columnwidth]{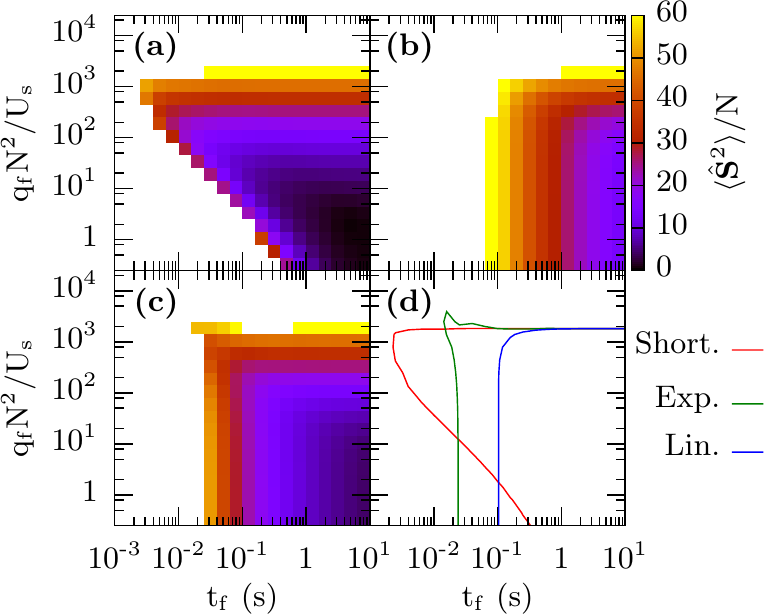}
\caption{Value of $\langle \hat{\bm S}^2 \rangle$ after the time evolution, 
i.e. $t=t_f$, for the three different protocols considered: shortcut (a), linear (b) 
and exponential (c). In all cases only the results that give 
$\langle \hat{\bm S}^2 \rangle < 60$ are shown. Finally, in (d), we show 
the contour lines corresponding to $\langle \hat{\bm S}^2 \rangle = 60$ . 
In all cases we have $N=500$ particles and an initial value of 
$q_0/h=0.1 \, U_s/h=10.413$ Hz. \label{fig:figmap1}}
\end{figure}

It is also interesting to see how the wave function evolves in time, going from 
a state with large $\langle \hat{\bm S}^2 \rangle$, where $c_S$ are centered around 
large $S$, to a state with small $\langle \hat{\bm S}^2 \rangle$, where the wave 
function takes substantial values around $S=0$ or $S=2$. In Fig.~\ref{fig:cs-3} we 
compare the wave functions at different times obtained 
with the shortcut (a,b) and exponential (c,d) protocols. The calculations correspond to 
the $N=100$ ones reported in Fig.~\ref{fig:fid-s2}. As can be clearly seen in 
all cases the wave function for the shortcut is much more peaked around $S=0$ 
than the exponential one. Also, as expected, the final wave function is more 
concentrated at smaller values of $S$ as we target final states with smaller $q_f$. 
This can be seen comparing panels (a,c), computed with $q_f N^2/U_s=0.01 N^2$ with 
panels (b,d), computed with $q_f N^2/U_s=0.001 N^2$. Finally, note that even though 
the shortcut performs quite well, the final wave function is peaked at $S=2$ rather than 
at $S=0$, which reflects the fact that we are still on the single BEC side of the crossover
reported  in Fig.~\ref{fig:cx} (c).

\section{Shortcut from BEC to a fragmented condensate}
\label{sec4}

\begin{figure}
\includegraphics[width=\linewidth]{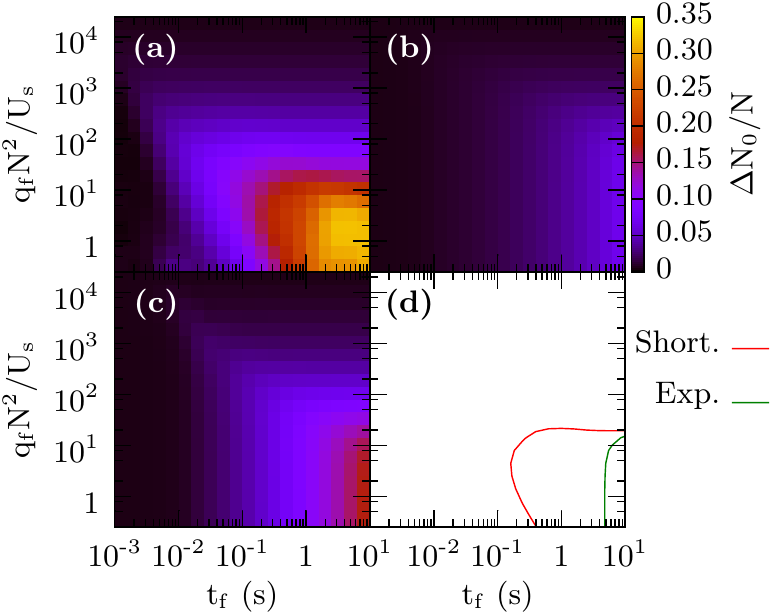}
\caption{$\Delta N_0 / N$ for the same conditions as in Fig.~\ref{fig:figmap1}. 
The results obtained from the three different protocols, the shortcut, 
a linear ramp and an exponential one are given in panels (a), (b) and (c), respectively. 
Finally, in (d) we show the contour lines of the three cases 
above for $\Delta N_0/N = 0.15$ . All these plots have been realized using a 
system of $N=500$ particles and $q_0/h=0.1 \, U_s/h=10.413$ Hz. \label{fig:figfra1}}
\end{figure}

In the previous section we have shown the superior performance of the shortcut 
protocol in comparison with exponential and linear ramps in 
the BEC regime, $qN^2/U_s\gg 1$. In this section we explore the fragmented condensate domain, that is, 
QZE ramps going from $q_0 N^2 /U_s\gg 1$ to $q_f N^2 /U_s \lesssim 1$. 

In Figs.~\ref{fig:figmap1} and \ref{fig:figfra1}, 
we provide an extensive comparison between our shortcut protocol and 
linear and exponential ramps. The figures depict the final values of 
$\langle \hat{\bm S}^2 \rangle$,  Fig.~\ref{fig:figmap1}, and
$\Delta N_0 / N$, Fig.~\ref{fig:figfra1}. In those figures we consider $N=500$ atoms, 
starting from the ground state corresponding to a value of $q_0 =0.1 U_s$. The 
figures cover a broad range of final target values of $q_f$ ranging from 
deep in the BEC sector into well below the transition to the fragmented 
condensate region, $q_f N^2/U_s \lesssim 1$, see Fig.~\ref{fig:cx} (Upper panel). 
Results are also reported as a function of the desired final time, $t_f$. 
We take again $U_s/h=104.13$ Hz and final times ranging from $0.001$ to $10$ seconds. 

As found previously, the shortcut protocol 
performs better than the exponential and linear ramps in the BEC region, 
as can be seen looking at the $q_f N^2 /U_s \gtrsim 1$ region in the three 
figures. For instance the region in the $(q_f N^2/U_s,t_f)$ map, where small 
final values of $\langle \hat{\bm S}^2 \rangle$ are larger for 
the shortcut protocol. The exponential produces also relatively low values, with a result 
mostly independent of the value of $q_f$, while the linear ramp fails to produce 
small final values, unless $t_f \simeq 10$ s.

In situations in which the target final state is clearly in the 
fragmented domain, $q_f N^2 /U_s\simeq 1$, the only method that 
produces sizeable fragmentation, as measured by $\Delta N_0 /N\gtrsim 0.15$, is the 
shortcut protocol (see Fig.~\ref{fig:figfra1}). The exponential ramp requires times 
almost two orders of magnitude larger to obtain the same level of 
fragmentation in the system. In line with the latter, lower final values 
for $\langle \hat{\bm S}^2 \rangle$ are obtained for the shortcut 
protocol for those cases in which the fragmentation is closer to the singlet 
value $\Delta N_0/N\simeq \sqrt{4/45}=0.298$~\cite{Ger1}. For the parameters 
considered here, a shortcut ramp performed in $t_f \sim 1 \,s$ is able to 
produce a state very close to the ground state, for $N=500$. Note that, using 
the notations and assumptions of Section~\ref{XXX}, the chosen value of 
$U_s/h \approx 10\,$Hz is achieved in a trap of frequency 
$\omega/(2\pi)\approx 300\,$Hz for $N=500$. The corresponding three-body 
lifetime is $\Gamma_3 \approx 1300\,$s$^{-1}$, or $N\Gamma_3 t_f \approx 0.4$: 
Less than a single three-body loss event (on average) during the entire 
shortcut protocol. Losses should not be a concern for $N\sim 500$.

\section{Comparison with current experimental setups}
\label{sec5}

\begin{figure}[t]
\includegraphics[width=0.9\linewidth]{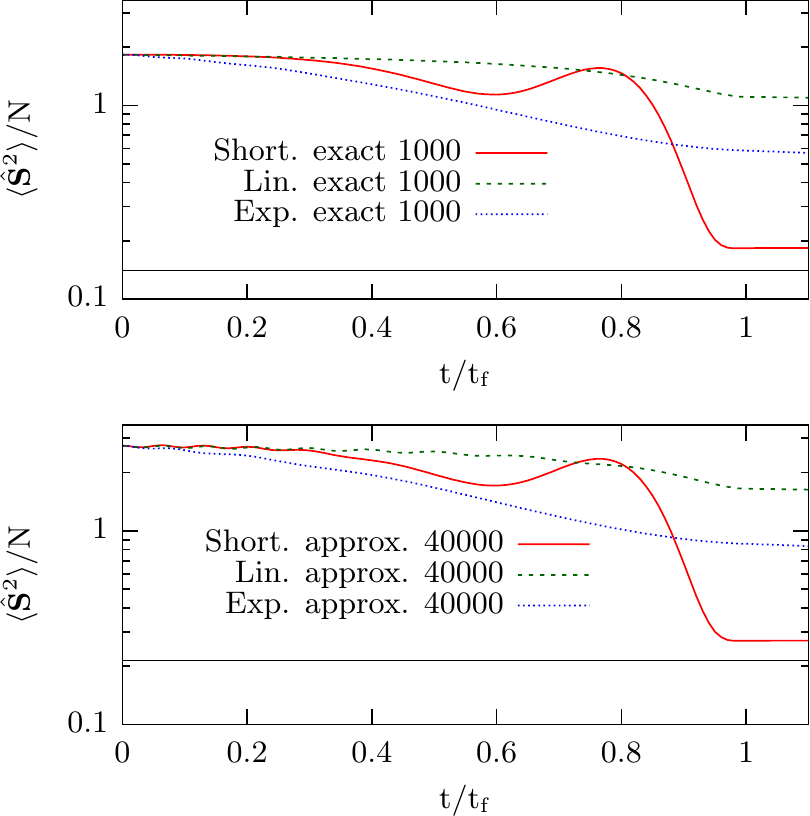}
\caption{$\langle \hat{\bm S}^2 \rangle$ as a function of time 
computed for systems evolved following the shortcut protocol, a linear or an 
exponential ramp. We compare together a system of $N = 1000$ spins (exact solution) 
and a system with $N=40000$ spins (approximate solution). The system is evolved 
from a state with $q_0/h=10\, U_s/h=71$ Hz to $q_f=10^{-3} q_0$ in $t_f=0.1$ s. The 
shortcut protocol also works for this set of parameters, although it is, in principle, 
only valid for $q(t) \ll U_s$, in the sense that it provides a clear gain over simpler 
exponential or linear ramps. \label{fig:comp}}
\end{figure}

We have been using, throughout the full manuscript, parameters taken from 
realistic proposals, most of them from~\cite{Ger1}. In this Section, we explore different 
parameters taken from other experimental setups. Some experiments~\cite{chap} have been 
recently done with $^{87}$Rb Bose condensates composed of $N=40000$ atoms with $U_s/h=7.1$ Hz and a $q(t)$ 
around $10\, U_s$. Taking these parameters, we have calculated the 
evolution of a system with $N=1000$ particles and $U_s/h=7.1$ Hz to 
check whether the shortcut protocol still gives good results under these 
experimental conditions. Results have also been obtained for a system with 
$N=40000$ spins using the shortcut protocol for the  approximate Hamiltonian 
in Eq.~(\ref{eq:ham3}). In Fig.~\ref{fig:comp} both results are shown for comparison.

$\langle \hat{\bm S}^2 \rangle$, shown in Fig.~\ref{fig:comp}, 
has been computed for these two systems and, although the initial values of 
$q(t)$ are larger than $U_s$, the shortcut protocol still drives the system 
to the ground state (or close) and improves the performance of the other two 
ramps. Although the shortcut protocol is, in principle, only valid for 
$q(t) \ll U_s$, this and other calculations (where we have driven a system 
from different values of $q_0$, all between 10 and 1000 times larger that 
$U_s$, to $q_f$ above and below $U_s$) show that the protocol can be successfully 
applied for larger $q_0$ values.

\section{Summary and conclusions}
\label{sec6}

We have presented a method to prepare a spin-1 BEC into a many-body spin singlet 
state by making use of an approximate protocol to shortcut the adiabatic following
in the many-body system. The protocol consists in specific functions $q(t)$ which 
are constructed such that the time evolution of the system brings the many-body 
state from the ground state for $q_0\equiv q(t=0)$ to the ground state for 
$q_f\equiv q(t=t_f)$. The main aim is to produce the very fragmented 
ground state of the spinor system in absence of quadratic magnetic field, starting 
from a condensate in the $m=0$ manifold in a regime dominated by the quadratic 
Zeeman term. The performance of the shortcut protocol has been compared to both 
a linear and an exponential ramp of the parameter $q$. 

Even though the protocol is only approximate, it is shown to provide a much better 
performance than the exponential and linear ones almost in all situations. In 
the BEC side, that is, for $qN^2/U_s \gg 1$, the method works almost perfectly for 
time intervals of the order of $1/U_s$ and larger. The method works also better 
for cases in which the BEC-Fragmented transition is targeted. In particular 
it works with similar accuracy as the exponential ramp up to times one 
order of magnitude smaller. To quantify the performance we have computed 
the achieved final value of $\langle \hat{\bm S}^2\rangle$ and the value of $\Delta N_0 /N$. 

We have obtained results for systems with different sizes and final and initial 
setups and we have seen that the protocol achieves better results for larger systems. 
Results have also been obtained from approximate solutions of the Schr\"odinger 
equation [using Eq.~(\ref{eq:ham3}) and~(\ref{eq:cx})]. Based on these results 
we have been able to extrapolate the method to larger systems and find that, 
with this protocol, a many-body spin singlet state can be obtained for many 
different systems sizes. We have also shown the success of our method when 
applied to systems prepared with parameters taken from current experimental 
setups. We believe that our method for preparing a BEC into a singlet state with 
short times is experimentally realizable and efficient. Further improvements to 
the shortcut protocol profiting from the available freedom inherent to the 
presented procedure will be the object of forthcoming investigations.

\acknowledgments
We acknowledge stimulating discussions with members of the Bose-Einstein condensates 
group at LKB, in particular with Bertrand Evrard and Jean Dalibard, and with Tommaso 
Roscilde. This work has been partially supported by DARPA (Optical Lattice 
Emulator Grant). We acknowledge financial support from the Spanish MINECO 
(FIS2014-54672-P), from Generalitat de Catalunya Grant No. 2014SGR401 and the Maria 
de Maeztu grant (MDM-2014-0369). LDS ackowledges support from the EU (IEF grant No. 236240) 
and TZ from the Hamburg Center for Ultrafast Imaging. B. J-D. is supported by the 
Ram\'on y Cajal MINECO program.


\appendix

\section{Eigenstates of the Hamiltonian in the $|N,S,M\rangle$ basis}
\label{app:A}

The Hamiltonian in Eq.~(\ref{eq:ham1}) 
is diagonalized by the total spin eigenstates $|N,S,M\rangle$, where $S$ is the total 
spin and $M$ the total projection of $S$ in the z-axis direction. A general state 
$\phi_M$ is written as, $|\phi_M \rangle=\sum_{S=|M|}^N c_S |N,S,M\rangle$. The 
construction of the angular momentum eigenstates is not trivial and these are built 
as follows~\cite{law1998a,ho2000a,koashi2000a,castin2001a,Ger1},
\begin{equation}
|N,S,M\rangle = \frac{1}{\mathcal{Z}(N,S,M)^{1/2}}\left(\hat S_-\right)^P
\left(\hat A^\dagger\right)^Q\left(\hat a_{+1}^\dagger\right)^S|{\rm vac}\rangle \, ,
\end{equation}
where $P=S-M$,  $2Q=N-S$, $\hat a_i^\dagger$ and $\hat a_i$ are the creation and 
annihilation operators of the state $i$ respectively, 
$\hat{S}_-=\sqrt{2}(\hat a_{-1}^\dagger \hat a_0 + \hat a_0^\dagger \hat a_{+1})$ is 
the lowering total spin operator and 
$\hat A^\dagger=(\hat a_0^\dagger)^2-2\hat a_{+1}^\dagger \hat a_{-1}^\dagger$ is the singlet creation operator.

The expression of these states involves three operators. The first one, $\hat a_{+1}^\dagger$ is 
the creation operator of a spin $1$ particle with $m=+1$. This operator acting $S$ times 
over the vacuum leads to the many-particle state $\propto |S,S,S\rangle$. The following acting 
operators $\hat A^\dagger$ and $\hat S_-$ commute with the total spin momentum 
operator and therefore do not modify $S$. The singlet creator operator creates pairs with total spin $0$, so it only 
changes the number of particles. Then, $Q$ repeated actions of this operator add 
singlet pairs until the state $\propto |N,S,S\rangle$ is obtained. Finally, the lowering angular 
momentum operator $S_-$ acts $P=S-M$ times without affecting $N$ or $S$, leading to the final 
state $\propto |N,S,M\rangle$. The normalization factor is obtained after tedious calculations,
\begin{equation}
\mathcal{Z}(N,S,M)=S!\ \frac{(N-S)!! (N+S+1)!!}{(2S+1)!!}\ \frac{(S-M)!(2S)!}{(S+M)!} \, . 
\end{equation}

The complete Hamiltonian in Eq.~(\ref{eq:ham1}) can be computed in the $N,S,M$ basis. The interaction 
term is diagonal, but the operator $\hat{N}_0=\hat a_0^\dagger \hat a_0$ has matrix elements between 
states with $S$ and $S \pm 2$. The total Hamiltonian is thus tridiagonal and therefore easy to solve numerically. 
The action of $\hat{N}_0$ is explicitly given by~\cite{Ger1},
\begin{eqnarray}
&&q\hat{N}_0 |N\ S\ M\rangle = \nonumber \\
&=&q\sqrt{A_-(N,S+2,M)A_+(N,S,M)}|N\ S+2\ M\rangle + \nonumber \\
&+& q\sqrt{A_+(N,S-2,M)A_-(N,S,M)} |N\ S-2\ M\rangle +  \nonumber \\
&+& q\left[ A_-(N,S,M)+A_+(N,S,M)\right] |N\ S\ M\rangle \, ,
\end{eqnarray}
where
\begin{eqnarray}
A_+(N,S,M)&=&\frac{(S+M+1)(S-M+1)(N-S)}{(2S+1)(2S+3)}\, , \nonumber\\
A_-(N,S,M)&=&\frac{(S+M)(S-M)(N+S+1)}{(2S+1)(2S-1)} \, . 
\end{eqnarray}
The resulting Hamiltonian, given also in Eq.~(\ref{eq:ham2}), reads, 
\begin{equation} 
h_{S,S+2}\, c_{S+2} +h_{S,S-2}\, c_{S-2} + h_{S,S}\, c_S= E\, c_S\, .
\end{equation}
with
\begin{eqnarray}\label{eq:h}
h_{S,S+2}&=&-q\sqrt{(N+S+3)(N-S)}  \nonumber \\
&\times& \frac{(S+1)(S+2)}{(2S+3)\sqrt{(2S+1)(2S+5)}}, \nonumber \\
h_{S,S-2}&=&-q\sqrt{(N+S+1)(N-S+2)} \\
&\times& \frac{S(S-1)}{(2S-1)\sqrt{(2S+1)(2S-3)}}, \nonumber \\
h_{S,S}&=&\frac{U_s}{2N}S(S+1)  \nonumber \\
&-& q\left[\frac{S^2(N+S+1)}{(2S-1)(2S+1)}+\frac{(S+1)^2(N-S)}{(2S+1)(2S+3)}\right] \, . \nonumber
\end{eqnarray}

\section{Continuum approximation of the Hamiltonian}
\label{app:B}

A continuum approximation of Eq.~(\ref{eq:ham2}) can be obtained by considering 
$1\ll S\ll N$. The wave function $c_S$ can, thus, be approximated by a continuous 
function $c(x)$, where $x\equiv S/N$ and varies from $0$ to $1$. Then, 
$\epsilon=2/N$ can be taken as a small parameter and a Taylor expansion can 
be made,
\begin{equation}
c_{S\pm 2} = c(x) \pm \epsilon c'(x) + \frac{\epsilon^2}{2} c''(x) + \mathcal{O}(\epsilon^3) \, .
\end{equation}
By substituting this expression into Eq.~(\ref{eq:ham2}) the following continuum 
Schr\"odinger equation is obtained
\begin{equation}
\alpha(x) c''(x) + \beta(x) c'(x) + \left(\gamma(x) - E\right) c(x) =0 \, , 
\end{equation}
where
\begin{eqnarray}
\alpha(x)&=&\frac{\epsilon^2}{2} (h_{S,S+2}+h_{S,S-2})\, ,\nonumber \\
\beta(x)&=&\epsilon (h_{S,S+2}-h_{S,S-2})\, ,\label{eq:abc} \\
\gamma(x)&=&h_{S,S}+h_{S,S+2}+h_{S,S-2} \nonumber \, .
\end{eqnarray}

Taking into account that $1\ll S\ll N$, we can set the order until which 
we want to approximate. Performing a Taylor expansion in $S/N$, $1/S$ and $1/N$, and 
substituting the resulting expressions in Eqs.~(\ref{eq:abc}), one finds
\begin{eqnarray}
\alpha(x)&\approx &-\frac{q}{N}\left(1-\frac{x^2}{2}+\frac{3}{2N}\right) \, , \nonumber \\ 
\beta(x)&\approx &\frac{-q}{4N^2 x^2}\left(1-\frac{x^2}{2}+\frac{3}{2N}\right)\, , \label{eq:abc2}
\\
\gamma(x)&\approx &\frac{N}{2}U_s x^2 -qN\left(1-\frac{x^2}{4}+\frac{1}{2N}+\frac{1}{8N^2 x^2}\right) \, . \nonumber
\end{eqnarray}
Keeping terms to leading order in $1/N, x$ we arrive at the Schr\"odinger-like 
equation~(\ref{eq:ham3}).

\end{document}